%% file: main.tex
\documentclass[conference]{IEEEtran}
\IEEEoverridecommandlockouts
\usepackage{cite}
\usepackage{amsmath,amssymb,amsfonts}
\usepackage{algorithmic}
\usepackage{graphicx}
\usepackage{textcomp}
\usepackage{xcolor}
\usepackage{multirow}
\usepackage{pifont}
\usepackage{bbding}
\usepackage{colortbl}

\usepackage{algorithm}
\usepackage{algorithmic}
\usepackage{booktabs}
\usepackage{enumitem}
\usepackage{url}

\def\BibTeX{{\rm B\kern-.05em{\sc i\kern-.025em b}\kern-.08em
    T\kern-.1667em\lower.7ex\hbox{E}\kern-.125emX}}
\begin{document}

\title{
{\huge [Experiment, Analysis, and Benchmark] }\\
CITBench: A Comprehensive Benchmark for Interactive Tabular Data Processing with LLMs\\
}

\author{\IEEEauthorblockN{Zihan Nan\IEEEauthorrefmark{1},
Yang Gu\IEEEauthorrefmark{1},
Wei Liu\IEEEauthorrefmark{1},
Xi Yan\IEEEauthorrefmark{2},
Zhou Liu\IEEEauthorrefmark{1},
Hao Liang\IEEEauthorrefmark{1}, and
Wentao Zhang\IEEEauthorrefmark{1}}

\IEEEauthorblockA{\IEEEauthorrefmark{1}Peking University, Beijing, China}
\IEEEauthorblockA{\IEEEauthorrefmark{2}Institute of Computing Technology, Chinese Academy of Sciences, Beijing, China}
\IEEEauthorblockA{\{zhnan25, zhouliu25, hao.liang, eularioal\}@stu.pku.edu.cn, \{gu\_yang, wentao.zhang\}@pku.edu.cn, yanxi25s@ict.ac.cn}
}

\maketitle

\begin{abstract}
Tabular data processing is central to data work, and LLM-based assistants have recently shown promising capabilities in supporting such tasks. However, existing benchmarks primarily focus on table reasoning under single-turn, fully specified instructions, underrepresenting complex table processing that unfolds through multi-turn interactions with evolving user requirements. To bridge this gap, we introduce CITBench, a comprehensive benchmark for evaluating LLMs on interactive tabular data processing. CITBench features a comprehensive taxonomy across four high-level categories—table matching, cleaning, augmentation, and transformation—spanning 18 task types and 1,296 instances curated from datasets across diverse domains. The benchmark supports both offline and online evaluation, where the online setting models multi-turn interactions under constrained operation procedures and structured task scripts, capturing key potential behavioral characteristics of user-in-the-loop tabular data processing. We evaluate a broad suite of open-source and closed-source LLMs on CITBench, revealing a consistent trend: while current models perform well on simple tables and rules, their performance degrades significantly with increasing table complexity, tighter rule dependencies, and noisy multi-turn interaction simulations. These results highlight persistent challenges in understanding, planning, and table-structure awareness for LLMs in extended interactive data processing scenarios. \footnote{The source data and implementation code are publicly available at \url{https://github.com/SSndot/CITBench}.}
\end{abstract}


\input{./intro.tex}
\input{./related.tex}
\input{./pre.tex}
\input{./CITBench.tex}
\input{./exp.tex}

\section{Conclusion}
In this paper, we introduce CITBench, a benchmark designed for a comprehensive evaluation of LLMs’ tabular data processing capabilities across representative task categories by simulating potential user interaction characteristics in multi-turn scenarios. Our evaluations reveal that current models exhibit a substantial performance gap between idealized single-shot settings and interactive environments: multi-turn task decomposition effectively unlocks models’ complex tabular processing capabilities, while cognitive noise inherent to simulated user interactions leads to universal performance degradation across all tested models. Future research should prioritize enhancing state tracking and intent resolution capabilities for tabular data processing, to equip agents with the robustness required to effectively handle persistent user perturbations in dynamic environments.

\bibliographystyle{IEEEtran}
\bibliography{refs}

\end{document}

%% file: intro.tex
\section{Introduction}
\begin{figure}[htbp]
  \vspace{-0.5em}
  \centering
  \includegraphics[width=\linewidth]{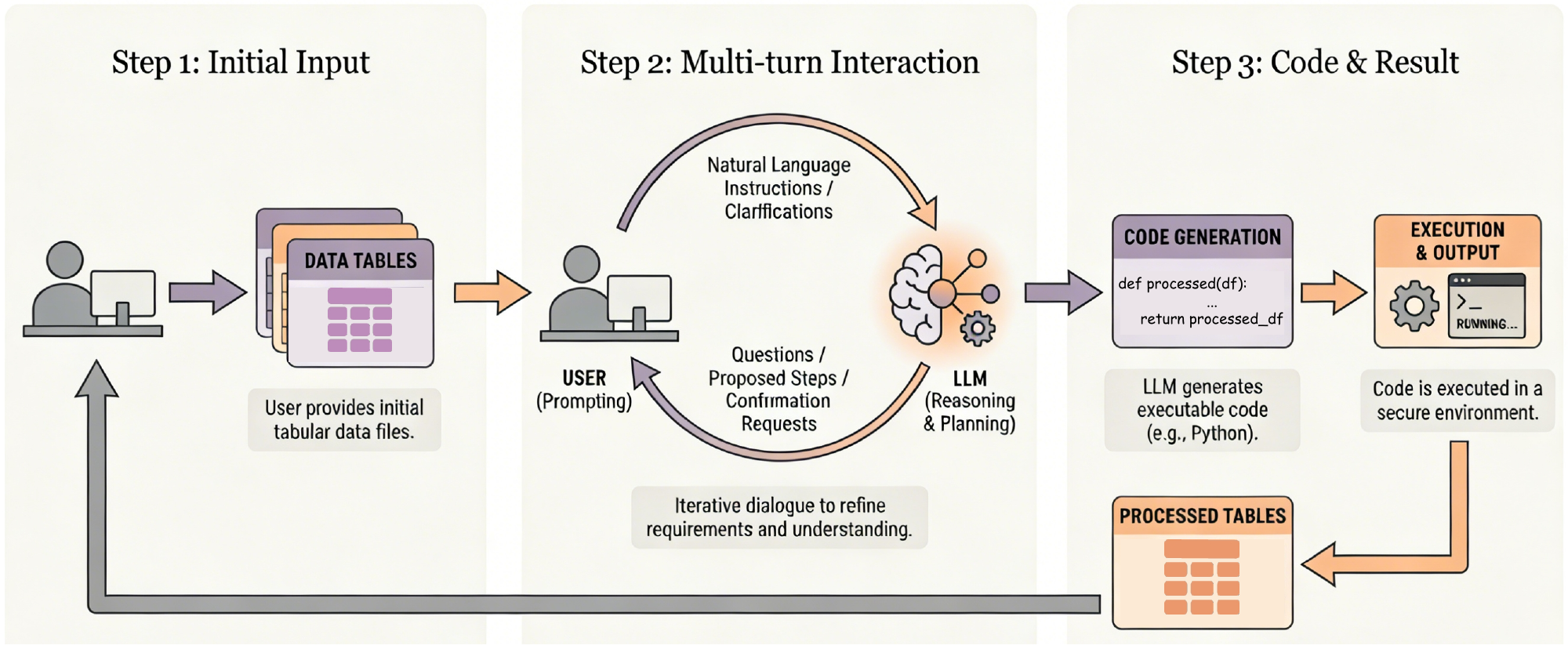}
  \caption{Tabular Data Processing via Multi-turn Interaction.}
  \label{case} 
  \vspace{-0.6em}
\end{figure}

Tabular data serves as the critical backbone for data-driven decision-making across diverse sectors, including financial analytics, healthcare informatics, e-commerce, and scientific research \cite{clements2020sequential, ulmer2020trust, rahman2020benchmarking}. In the contemporary era of data-centric AI, the analytical value of these datasets is fundamentally predicated on effective tabular data processing—sophisticated workflows dedicated to transforming raw, misaligned, and frequently noisy tables into high-quality, actionable knowledge assets \cite{jain2020overview, jassim2021data, lu2025large}. Recently, large language model (LLM)-based methods have demonstrated promising baseline results across a wide range of atomic data engineering operations, including schema-level table matching \cite{zhang2023jellyfish, xing2025table}, systematic data cleaning \cite{ni2024iterclean, huang2024cocoon, naeem2024retclean, 11113193}, row/column augmentation \cite{sui2024tap4llm, li2024table}, and schema-preserving transformations \cite{zhang2025tablellm, qian2024unidm}. However, despite these algorithmic advances, the practical performance of current systems remains severely bounded by an idealized evaluation assumption: that users can provide complete, unambiguous, and static requirements within a single shot. In practice, data engineering tasks are inherently collaborative and iterative, unfolding through multi-round conversational exchanges where instructions are initially underspecified or fluid, as illustrated in Fig.~\ref{case}. This dynamic interaction introduces pervasive operational noise—stemming from ambiguous descriptions, incomplete rule propagation, and frequent requirement revisions \cite{zhang2024clamber, herlihy2024overcoming, zamfirescu2023johnny, 10598154}—presenting distinct cognitive and planning challenges that current single-pass evaluation paradigms completely fail to capture.

\begin{table*}[!hpt]
    \caption{Comparison of representative tabular data processing benchmarks across task categories and task dimensions.}
    \vspace{-0.8em}
    \label{tab-BenchCompare}
    \centering
    \renewcommand{\arraystretch}{1.05}  
    \setlength{\tabcolsep}{1.8mm}{
    \footnotesize
    \begin{tabular}{c|cccc|ccc}
    \hline
    \multirow{2}{*}{\textbf{\fontsize{8pt}{\baselineskip}\selectfont  Benchmark}} & \multicolumn{4}{c|}{\textbf{Task Category}}                                                                     & \multicolumn{3}{c}{\textbf{Task Dimension}}                                       \\ \cline{2-8} 
            & \textbf{\begin{tabular}[c]{@{}c@{}}Table\\ Matching\end{tabular}} & \textbf{\begin{tabular}[c]{@{}c@{}}Table\\ Cleaning\end{tabular}} & \textbf{\begin{tabular}[c]{@{}c@{}}Table\\ Augmentation\end{tabular}} & \textbf{\begin{tabular}[c]{@{}c@{}}Table\\ Transformation\end{tabular}} & \textbf{\begin{tabular}[c]{@{}c@{}}Data\\ Scope\end{tabular}} & \textbf{\begin{tabular}[c]{@{}c@{}}Operational\\ Complexity\end{tabular}} & \textbf{\begin{tabular}[c]{@{}c@{}}Interaction\\ Mode\end{tabular}} \\ \hline
    \textbf{SpreedsheetBench} \cite{ma2024spreadsheetbench}    & \ding{53}  & \ding{53}  & \ding{53}  & \Checkmark  & Single\&Multi-table  & Single-step  & Offline  \\
    \textbf{SheetRM} \cite{chen2025sheetagent}    & \ding{53}  & \ding{53}  & \ding{53}  & \Checkmark  & Single\&Multi-table  & Multi-step  & Offline  \\
    \textbf{AutoDCWorkflow} \cite{li2024autodcworkflow}   & \ding{53}  & \Checkmark   & \ding{53}  & \ding{53}  & Single-table  & Multi-step  & Offline  \\
    \textbf{DataGovBench} \cite{liu2025datagovbench}   & \Checkmark  & \Checkmark  & \ding{53}  & \Checkmark  & Single\&Multi-table  & Single\&Multi-step   & Offline  \\
    \cellcolor{gray!20} \textbf{CITBench(ours)}    & \cellcolor{gray!20} \Checkmark   & \cellcolor{gray!20} \Checkmark    & \cellcolor{gray!20} \Checkmark   & \cellcolor{gray!20} \Checkmark   & \cellcolor{gray!20} Single\&Multi-table & \cellcolor{gray!20} Single\&Multi-step   & \cellcolor{gray!20} Offline\&Online  \\ \hline
    \end{tabular}
    }
\end{table*}

To address these gaps, establishing a comprehensive and rigorous benchmark for multi-turn tabular data processing is essential. As systematically compared in Table~\ref{tab-BenchCompare}, existing evaluation frameworks exhibit a historical disconnect between cell perception and active data manipulation. Traditional tabular benchmarks primarily focus on static, read-only table reasoning such as question answering or relational fact verification \cite{pasupat2015compositional, chen2019tabfact}, leaving the models' active cell-matrix transformation capabilities unassessed. More recent processing-oriented benchmarks \cite{ma2024spreadsheetbench, li2023sheetcopilot, li2024autodcworkflow} take meaningful steps toward executable table operations by requiring python script or spreadsheet macro generation; however, they remain confined to isolated, single-turn executions and largely overlook the iterative, user-in-the-loop nature of comprehensive data workflows \cite{kandel2011wrangler}, where requirements are progressively clarified through continuous feedback rather than prescribed upfront. Furthermore, while general-purpose conversational agent frameworks \cite{yang2023intercode, lai2023ds} provide a baseline for multi-turn execution, they lack the domain-specific depth required to enforce relational cell constraints or track structural table states under conversational noise. Consequently, existing frameworks fall short of stress-testing autonomous data assistants under interactive, long-horizon data processing scenarios.

To bridge these gaps, we introduce \textbf{CITBench}, a comprehensive benchmark for evaluating LLMs in interactive tabular data processing settings. CITBench establishes a unified task taxonomy spanning four high-level domains—table matching, cleaning, augmentation, and transformation—further partitioned into 18 distinct task types and 1,296 benchmark instances curated from diverse publicly available datasets across 10 industrial domains. The benchmark operates via a robust dual-stream construction framework. The \textbf{\emph{offline}} pipeline first synthesizes high-quality, verifiable task specifications through a template-driven synthesis-and-verification engine across multiple difficulty tiers. The \textbf{\emph{online}} pipeline decomposes these specifications into dependency-preserving subtask sequences to construct multi-turn interactive trajectories. 

Crucially, to support scalable and reproducible evaluation of interactive behaviors without unscalable manual logs, CITBench incorporates a Cognitive Simulator equipped with two primary mechanisms: the \emph{Cognitive-Load Window} and the \emph{Perturbation-Resolution Cycle}. Informed by our controlled user study of 10 participants, these components model bounded human cognitive capacities by limiting the per-turn instruction subtasks, while concurrently injecting structured conversational noise (such as fuzzy instructions, rule distortions, and redundancies) alongside next-turn resolution queues. This interactive design forces the model to continuously track table states and reconcile requirement updates across non-linear execution trajectories.

Extensive evaluations on CITBench across 13 state-of-the-art open-source and closed-source LLMs spanning six major architectural families reveal a profound fragility in current models. While top-tier closed-source systems demonstrate competitive precision in single-turn, fully specified settings, the injection of simulated conversational noise and multi-step rule dependencies triggers widespread planning failures and context drift across all model scales. This vulnerability intensifies in the online configuration, where cognitive noise induces substantial precision drops, highlighting critical limitations in current models' contextual resilience and validating CITBench as a non-trivial diagnostic stress-test for next-generation data assistants.

Our main contributions are summarized into the following three pillars:
\begin{itemize}
\item[$\bullet \,$] \textbf{Structured Task Taxonomy}: We present a unified, granular taxonomy of tabular data processing tasks mapping 18 distinct operation types across table matching, cleaning, augmentation, and transformation, establishing a structured foundation for autonomous data engineering.
\item[$\bullet \,$] \textbf{Interactive Simulation Framework}: We propose CITBench, an interaction-driven benchmark powered by an offline synthesis pipeline and an online Cognitive Simulator, enabling systematic and scalable stress-testing of multi-turn planning, error reconciliation, and state-tracking capabilities under controlled conversational noise.
\item[$\bullet \,$] \textbf{Comprehensive Empirical Evaluation}: We conduct extensive benchmarking on 13 open-source and closed-source models spanning six prominent families, uncovering key capability boundaries and analyzing the core architectural factors influencing interactive table-processing performance across six proposed evaluation dimensions.
\end{itemize}

\begin{figure*}[htbp]
  \centering
  \includegraphics[width=0.95\textwidth]{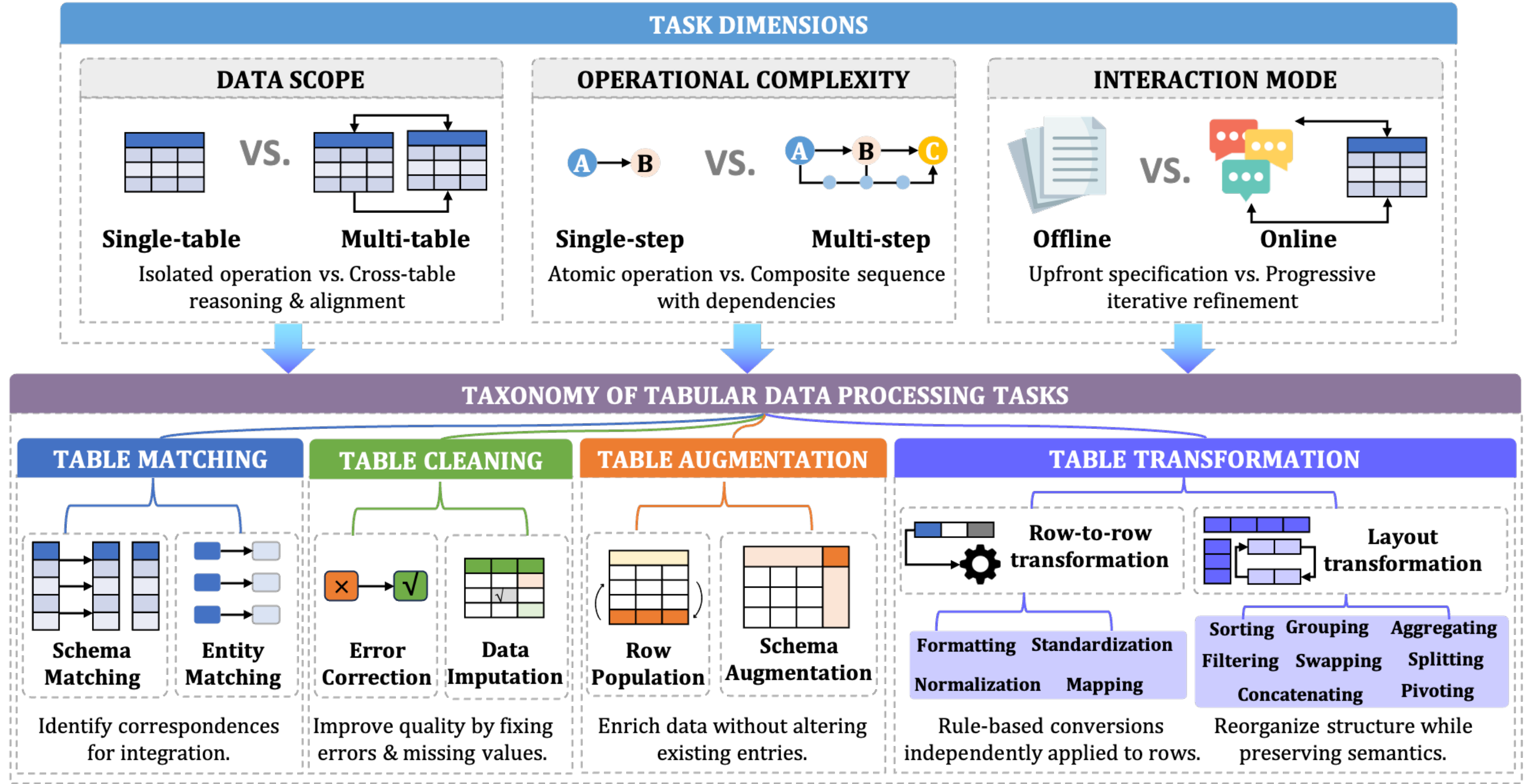}
  \vspace{-0.5em}
  \caption{Taxonomy and Dimensions of CITBench.}
  \label{fig-TaxDim} 
\end{figure*}

%% file: related.tex
\section{Related Work}


\subsection{Tabular Data Processing Benchmarks}

Unlike traditional tabular benchmarks that primarily focus on read-only reasoning, such as question answering and fact verification \cite{wu2025tablebench, wu2025realhitbench}, tabular data processing benchmarks evaluate LLMs’ ability to actively manipulate and transform data through executable artifacts, including code, formulas, or API calls. Early works such as SpreadsheetCoder \cite{chen2021spreadsheetcoder} and InstructExcel \cite{payan2023instructexcel} focus on formula prediction and command-to-operation mapping in spreadsheet environments. More recent benchmarks consider more realistic manipulation scenarios. SpreadsheetBench \cite{ma2024spreadsheetbench} collects real-world problems from Excel forums to evaluate robustness under diverse requests, while SheetRM \cite{chen2025sheetagent} emphasizes reasoning-dependent operations with long-horizon goals. In the domain of data quality and governance, AutoDCWorkflow \cite{li2024autodcworkflow} targets automated data cleaning workflows, and DataGovBench \cite{liu2025datagovbench} introduces a hierarchical evaluation framework for tasks such as deduplication and data integration.

Despite these advances, existing benchmarks are often limited to specific operation types or narrow subsets of tabular processing tasks. CITBench complements prior work by providing a more comprehensive benchmark that spans four categories of tabular data processing tasks, enabling broader evaluation beyond task-specific benchmarks (see Table~\ref{tab-BenchCompare}).

\subsection{Online and Multi-turn Interaction Benchmarks}

The evaluation of LLM agents increasingly moves from static instruction following toward online and multi-turn interaction benchmarks that emphasize decision-making under feedback and evolving contexts \cite{yao2024tau}. Early general-purpose benchmarks, such as MT-Bench \cite{zheng2023judging}, focus on multi-turn dialogue quality with limited environmental grounding, while DynaSaur \cite{nguyen2024dynasaur} incorporates execution feedback in interactive settings. More recent benchmarks ground interaction in executable environments. CodeAssistBench \cite{kim2025codeassistbench} evaluates multi-turn programming assistance with iterative execution feedback, and IDA-Bench \cite{li2025ida} studies conversational workflows for interactive data analysis. Although these benchmarks extend beyond single-turn evaluation, they typically assume stable goals and rational user behaviors, resulting in linear and idealized interaction trajectories. Moreover, they are not designed for structured tabular data processing tasks, which prior data management and spreadsheet systems characterize as inherently iterative and user-in-the-loop \cite{kandel2011wrangler}. As a result, interaction patterns specific to table-centric workflows remain underexplored in existing online interaction benchmarks.

CITBench is the first benchmark that systematically evaluates online and multi-turn interactions specifically for tabular data processing agents, addressing interaction requirements that are central to real-world table processing scenarios with evolving user intent and incomplete context.

%% file: pre.tex
\section{Preliminary}
\label{Preliminary}

\subsection{Taxonomy of Tabular Data Processing Tasks}


We organize tabular data processing tasks into a unified taxonomy that reflects common operations in real-world data management workflows. The taxonomy, as shown in Fig. \ref{fig-TaxDim}, consists of four high-level categories—\emph{table matching}, \emph{table cleaning}, \emph{table augmentation}, and \emph{table transformation}—each further divided into representative subcategories that capture distinct processing objectives and execution patterns.

\begin{itemize}

\item[$\bullet \,$] \textbf{Table Matching} focuses on identifying correspondences across tables to enable integration. It includes \textbf{\emph{schema matching}}, which aligns semantically equivalent columns based on predefined rules or constraints, and \textbf{\emph{entity matching}}, which identifies rows referring to the same real-world entities using key attributes. These tasks typically operate on multiple tables and require reasoning over both schema-level and instance-level information.

\item[$\bullet \,$] \textbf{Table Cleaning} aims to improve data quality by detecting and correcting errors within a table. This category includes \textbf{\emph{error correction}}, which fixes invalid or inconsistent values according to specified rules, and \textbf{\emph{data imputation}}, which fills missing values based on contextual or statistical requirements. Cleaning tasks generally preserve the table structure while modifying existing cell values.

\item[$\bullet \,$] \textbf{Table Augmentation} enriches tabular data by adding new information without altering existing entries. It consists of \textbf{\emph{row population}}, which inserts new rows based on external data or partial specifications, and \textbf{\emph{schema augmentation}}, which introduces additional columns derived from given rules or features. These tasks expand the table along either the row or column dimension.

\item[$\bullet \,$] \textbf{Table Transformation} modifies the representation or organization of tabular data to support downstream analysis. It includes two major forms: \textbf{\emph{row-to-row transformation}}, which apply rule-based conversions independently to individual rows (e.g., formatting, standardization, normalization, or value mapping), and \textbf{\emph{layout transformation}}, which reorganize the table structure through operations such as sorting, grouping, aggregation, filtering, or reshaping. While diverse in form, these transformations share the goal of restructuring data while preserving its analytical semantics.
\end{itemize}

This taxonomy provides a structured view of tabular data processing tasks and serves as the foundation for task instantiation and benchmark design in CITBench. Fine-grained task variants are derived from these categories to reflect practical data processing scenarios.

\subsection{Task Dimensions in CITBench}
Beyond task taxonomy, CITBench characterizes tabular data processing tasks along three orthogonal dimensions that reflect practical variations in data scope, operational complexity, and interaction mode.

\begin{itemize}

\item[$\bullet \,$] \textbf{Single-table vs.\ Multi-table tasks.}
This dimension distinguishes tasks based on whether they involve a single table or multiple related tables. Single-table tasks operate on an isolated table without cross-table reasoning. In contrast, multi-table tasks involve two or more tables with explicit relational dependencies, such as primary–foreign key relationships or shared entity identifiers. These tasks require agents to identify table roles (e.g., main and auxiliary tables) and perform operations that depend on cross-table alignment and integration, rather than simply processing multiple inputs.

\item[$\bullet \,$] \textbf{Single-step vs.\ Multi-step tasks.}
This dimension captures the complexity of the operator sequence required to complete a task. Single-step tasks correspond to an atomic tabular processing operation instantiated from a fine-grained task type defined in the taxonomy (e.g., a single formatting, filtering, or schema matching operation). Multi-step tasks consist of multiple heterogeneous atomic operations composed into an ordered sequence, with intermediate dependencies that may arise across steps. This distinction reflects the difference between isolated operations and composite workflows in real-world data processing.

\item[$\bullet \,$] \textbf{Offline vs.\ Online tasks.}
This dimension differentiates tasks by their interaction mode. Offline tasks assume that users provide a complete and explicit specification upfront, enabling execution in a single pass. Online tasks involve multi-turn interactions in which task specifications are progressively refined through user feedback. In such settings, user instructions may be initially ambiguous or partially incorrect and are adjusted over successive interaction rounds. CITBench models these interaction characteristics to reflect realistic user behaviors, which are further formalized in the cognitive simulation framework described in Section~\ref{CITBench}.

\end{itemize}

\begin{table}[htbp]
\caption{Examples of Extracted Instructional Noise Profiles from the User Study.}
\label{tab-user-study-noise}
\centering
\renewcommand{\arraystretch}{1.1}
\setlength{\tabcolsep}{2.5mm}{
\footnotesize
\begin{tabular}{l|l}
\hline
\textbf{Noise Profile} & \textbf{User Log Snippet (Example)} \\ \hline
\textbf{N1. Fuzzy} & \textit{"Format the date column to a unified form."} \\ 
\textbf{N2. Biased} & \textit{"Convert ages into 3 groups... wait, I mean 4."} \\ 
\textbf{N3. Redundant} & \textit{"Plot a bar chart before filling missing values."} \\ 
\textbf{N4. Order Swapped} & \textit{"Wait, swap the order: filter before grouping."} \\ \hline
\end{tabular}
}
\end{table}

\subsection{Empirical Analysis of User Interaction Patterns}

To better characterize the interaction dynamics of online tasks, we conduct a controlled user study on user-in-the-loop table manipulation. Unlike offline settings, where task specifications are fully provided upfront, online tasks unfold through multi-turn interactions in which users refine their intent, correct earlier instructions, and respond to intermediate system outputs. To capture these realistic patterns, we recruit 10 participants with data-science backgrounds and ask them to complete complex, multi-step tabular workflows using state-of-the-art conversational tabular AI systems. By auditing and formalizing the annotated interaction logs, we identify two core high-level cognitive action processes and four prevalent instructional noise profiles, which together inform the interaction modeling in CITBench.

\begin{enumerate}
    \item \textbf{Fuzzification--Refinement Process}: Users inherently possess a bounded cognitive load capacity when formulating data specifications. Instead of presenting exhaustive, deterministic rule parameters in a single shot, human users consistently demonstrate a progressive elicitation habit. They initiate interactions with broader, semantic-level intents, and only introduce granular operational boundaries (such as formatting syntax or cell thresholds) across subsequent refinement turns as the dialogue deepens.
    \item \textbf{Deviation--Correction Process}: Because tabular AI platforms visually present the processed cell matrices in real-time, users frequently engage in an immediate validation cycle. Whenever a cognitive misalignment occurs—wherein the user's expressed prompt inadvertently deviates from their latent objective—the user instantly recognizes the structural or cell-content discrepancy in the model's intermediate output. Consequently, they issue immediate modification or reordering commands in the next turn to correct the trajectory.
\end{enumerate}

Our empirical log analysis further formalizes these transactional friction patterns into four distinct linguistic and operational noise profiles, for which representative examples from our annotated logs are compiled in Table~\ref{tab-user-study-noise}: \textit{Fuzzy Instructions} (intentional omission of cell-level operational constraints), \textit{Biased Rules} (accidental distortion of mapping rules), \textit{Redundancy Noise} (insertion of standalone, non-blocking auxiliary subtasks such as temporary statistical queries), and \textit{Order Perturbations} (cognitive sequence swaps between dependent table operators). These extracted behavioral dynamics and structural noise distributions provide the objective empirical grounding for our interactive simulation architecture detailed in Section~\ref{CITBench}.

%% file: citbench.tex
\section{CITBench}
\label{CITBench}

\begin{figure}[htbp]
  \centering
  \includegraphics[width=0.47\textwidth]{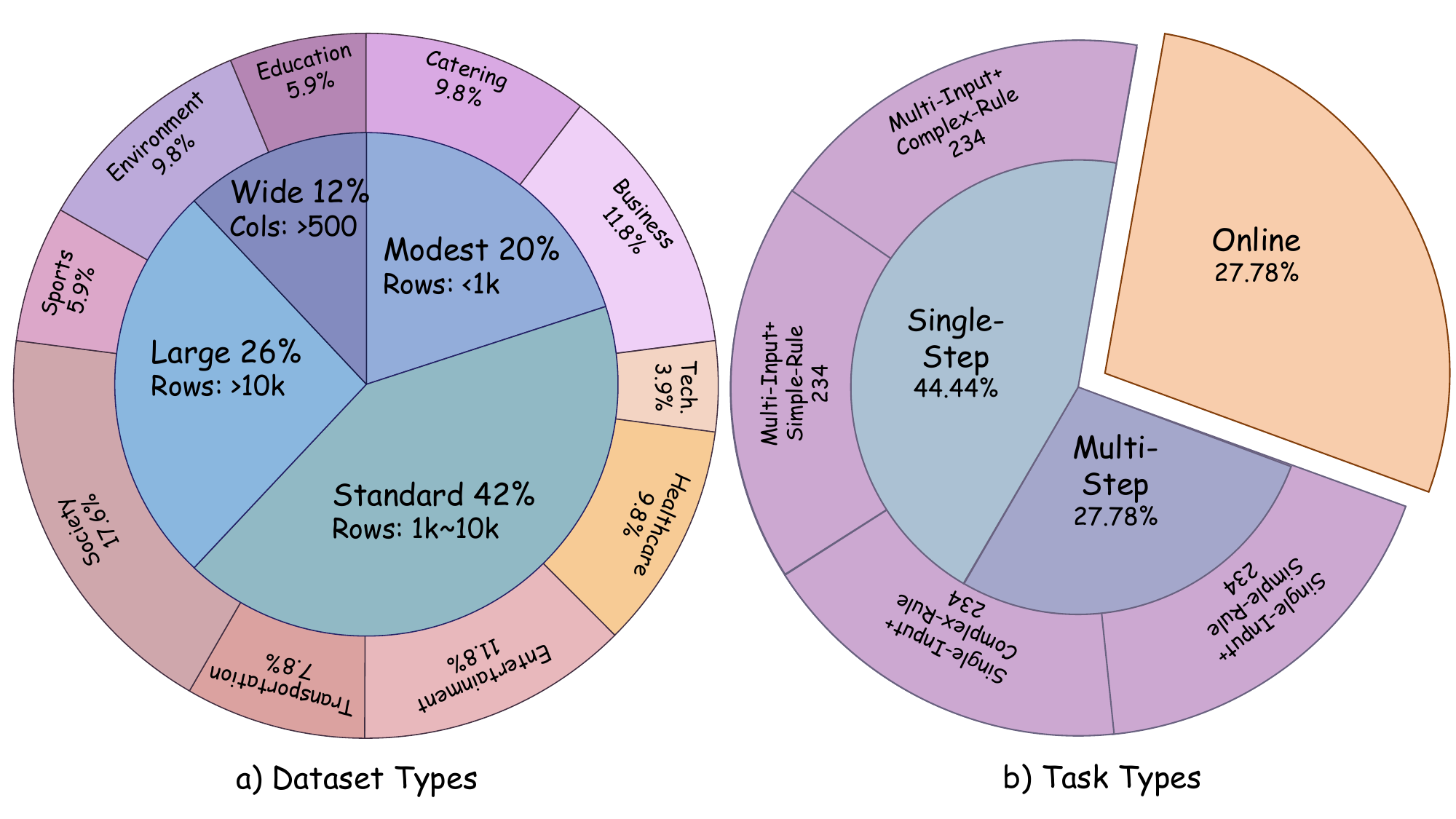}
  \vspace{-0.5em}
  \caption{Key statistics of CITBench.}
  \label{type} 
  \vspace{-0.5em}
\end{figure}

This section details the construction of CITBench. As shown in Fig.~\ref{type}, CITBench is designed to capture realistic tabular intelligence scenarios through diverse task compositions, covering both offline and online settings as well as single-step, multi-step, and multi-input tasks under simple and complex rules. Meanwhile, it is built on diverse table types with varying scales and domains.

\begin{figure*}[htbp]
  \centering
  \includegraphics[width=0.85\textwidth]{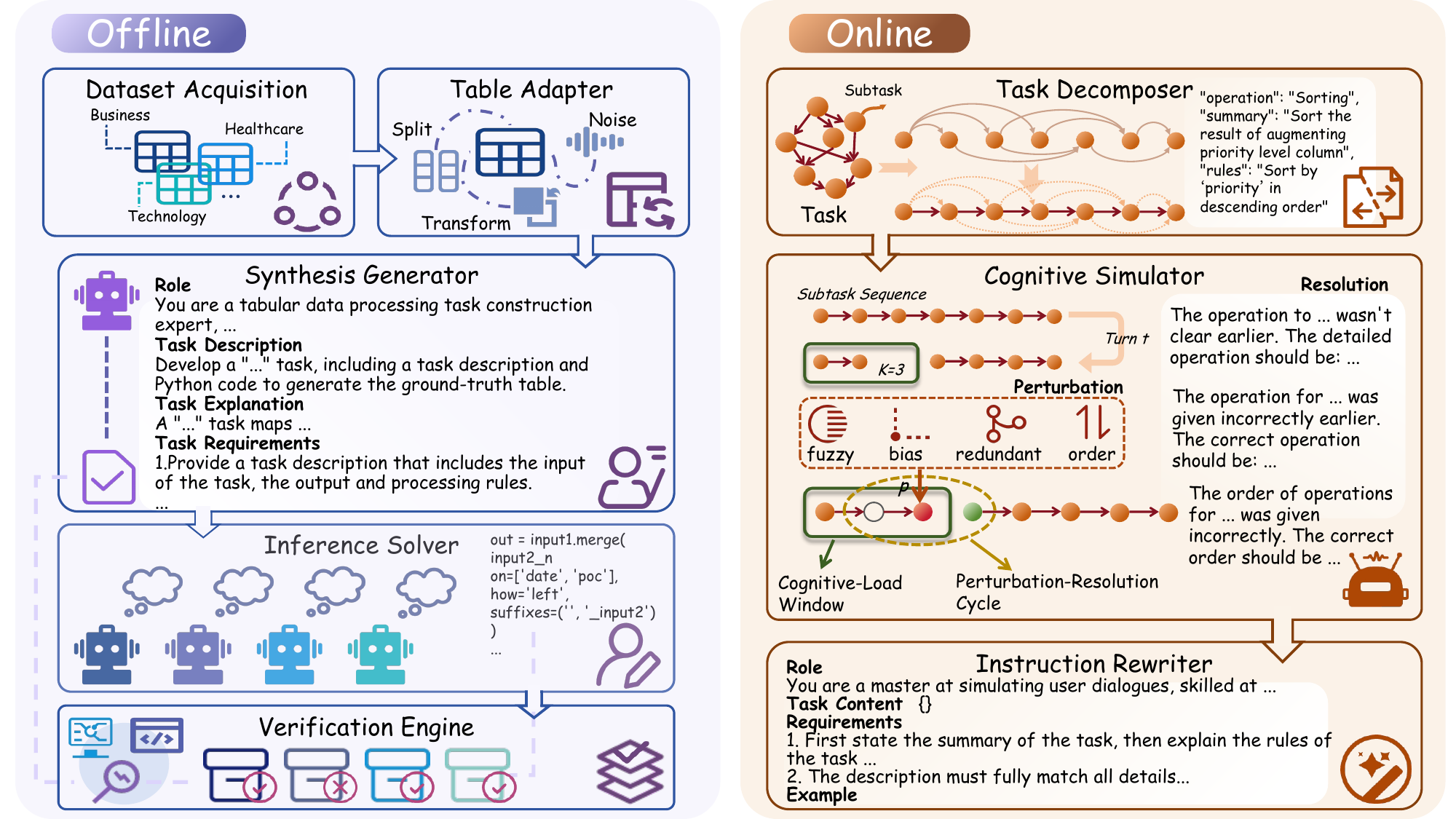}
  \caption{Overview of the Offline and Online task construction pipelines in CITBench.}
  \label{pipeline} 
\end{figure*}

To systematically construct such a diverse and reliable benchmark, we organize CITBench around three orthogonal core pillars: \textbf{Offline Task Synthesis} (the generative base), \textbf{Quality Assurance and Verification} (the reliability lock), and \textbf{Online Interaction Simulation} (the multi-turn dynamic protocol). Fig.~\ref{pipeline} illustrates the end-to-end task generation and validation pipeline.

\subsection{Offline Task Synthesis}
The offline stratum forms the baseline dataset pool of CITBench, dedicated to curating foundational multi-step data processing structures across heterogeneous matrices.

\paragraph{Data Acquisition and Table Adaptation}
We curate 50 public source datasets across 10 distinct domains (e.g., Catering, Business, Technology, and Healthcare), meticulously balanced across four dimensional table regimes (Modest, Standard, Large, Wide) as characterized in Fig.~\ref{type}. To transform these raw multi-domain source matrices into task-aligned data evaluation assets, the {\itshape Table Adapter} engine implements three primitive structural operations: 1) {\itshape Split} partitions source tables by rows or columns while maintaining primary-foreign or shared entity linkage keys; 2) {\itshape Noise} probabilistically injects data quality disturbances (missing value patterns, structural outliers, header synonyms) to simulate messy data states; and 3) {\itshape Transform} reorganizes tables into linkable multi-table views.

\paragraph{Candidate Specification Generation}
The adapted matrices and rule templates are routed to the {\itshape Synthesis Generator} (instantiated via Claude-Sonnet-4.5) to produce candidate execution instructions and corresponding Python reference snippets. The generation process enforces a unique result mapping tailored to 32 explicit difficulty categories defined by rule complexity, subtask diversity, table multiplicity, and matrix scale.

\subsection{Quality Assurance and Verification Framework}
To eliminate model-contingent filtering bias and guarantee that CITBench measures generalizable tabular reasoning rather than trivial heuristics, we introduce a rigorous automated-and-human joint verification framework.

\paragraph{Cross-Model Consensus Filtering}
To establish non-trivial difficulty boundaries while mechanically validating answer determinism, candidate tasks generated by the {\itshape Synthesis Generator} are evaluated by six leading code-capable models acting as independent {\itshape Inference Solvers} (Claude-Opus-4.5, Claude-Sonnet-4.5, Gemini-3-Pro-Preview, GPT 5.2, GLM 4.7, and DeepSeek V3.2). We enforce a strict cross-model consensus filter: a task and its ground-truth cell matrix are accepted into the raw benchmark pool if and only if more than one-third ($>1/3$) of these independent solvers yield identical target tables, while ensuring that at least one model fails. 

The mathematical rationality of this threshold design is centered on simultaneous correctness verification and difficulty preservation. By demanding consensus among independent multi-family architectures, the filter guarantees ground-truth correctness while preventing the inclusion of trivial tasks where full cross-model success occurs. To eliminate potential correctness bias, this heuristic is mathematically bounded by strict structural templates that enforce unique answer mappings, and is supplemented by an absolute human verification gate. Furthermore, representativeness bias is completely averted as the macro-distribution of task types remains structurally fixed by predefined taxonomic dimensions rather than fluctuating based on model performance. \textbf{An empirical analysis of tasks excluded by the $>1/3$ consensus rule demonstrates that these rejections stem entirely from minor instruction-level layout or description ambiguities}—where distinct models produce divergent but mathematically justifiable matrix variations—rather than model-specific capability gaps, confirming that our consensus protocol isolates objective description quality rather than introducing model-contingent filtering artifacts.

\paragraph{Sandbox Execution and Attribution Engine}
The automated {\itshape Verification Engine} provides an isolated, fully dependency-equipped sandbox environment to compile and execute reference code. It automatically intersects the outputs of the {\itshape Inference Solvers} with the generator's targets. Concurrently, it captures all failure signals—such as syntax faults, runtime errors, and matrix structural mismatches—and automatically attributes fine-grained root causes to predefined failure categories for systematic model debugging analysis.

\paragraph{Two-Stage Expert Human Validation Protocol}
As the ultimate quality gatekeeper to guarantee 100\% ground-truth target precision, we recruit 3 expert computer science and data science graduate students with rich programming backgrounds to conduct a double-pass manual validation protocol:
\begin{itemize}[leftmargin=12pt]
    \item \textit{Stage 1: Task Rationality Verification.} Validators independently audit all synthesized text instruction prompts and constraints to eliminate structural and semantic ambiguity, ensuring that each problem formulation contains a mathematically complete set of conditions to guarantee unique answer convergence. A logging analysis of tasks excluded at this stage confirms that rejections stem entirely from minor instruction layout ambiguities (where different interpretation heuristics yield multiple reasonable cell variants) rather than model capability dropouts, validating baseline distribution independence.
    \item \textit{Stage 2: Ground-Truth Re-verification.} Validators perform exhaustive, cell-by-cell matrix comparisons against the formalized requirements. Every target table is cross-checked to verify that operations (e.g., matching linkages, data cleaning imputations, or schema-preserving transformations) are flawlessly executed, securing absolute data integrity before sample locking.
\end{itemize}

\begin{figure}[htbp]
  \centering
  \includegraphics[width=0.47\textwidth]{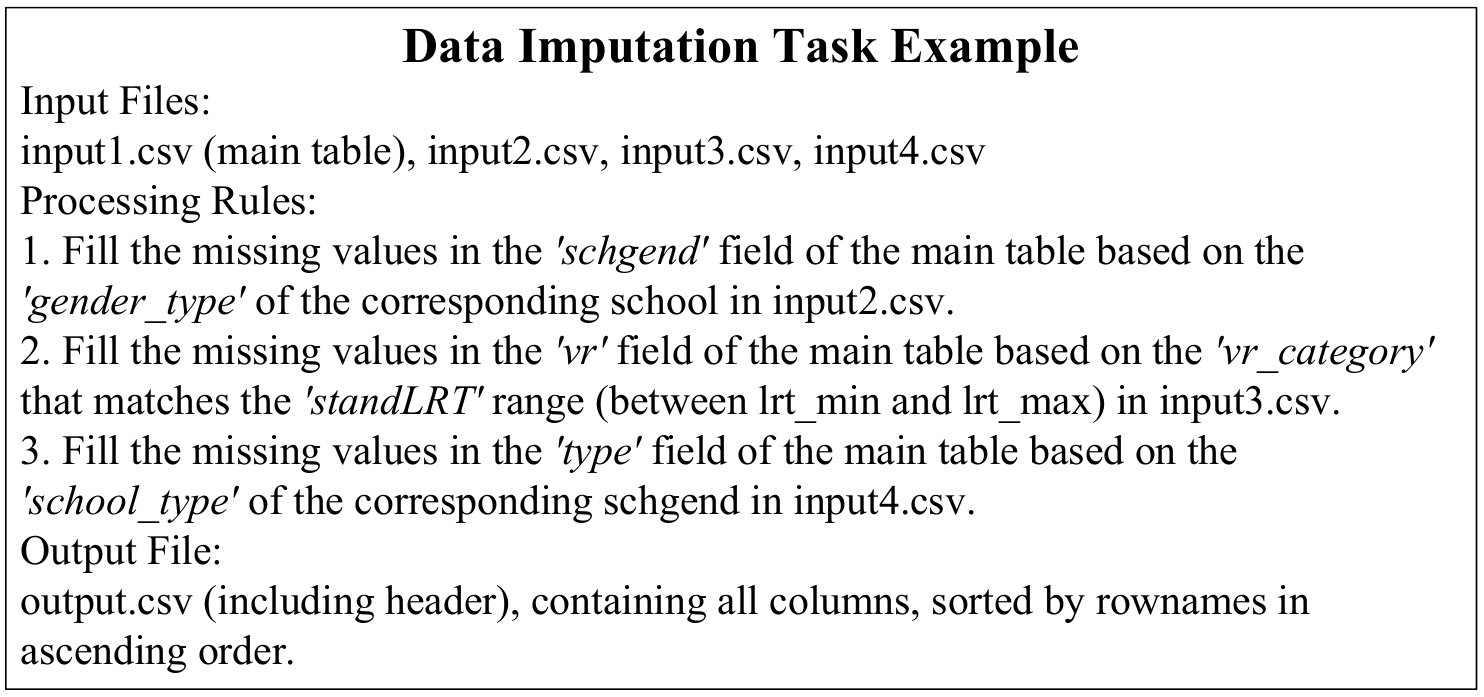}
  \vspace{-0.5em}
  \caption{An example task specification for the Data Imputation category in offline tasks.}
  \label{task} 
  \vspace{-0.5em}
\end{figure}

\subsection{Online Interaction Simulation}
The online configuration models the collaborative, multi-turn conversational trajectories of practical data work, translating the static multi-step offline specifications into sequential, interactive evaluation environments.

\paragraph{Task Decomposer}
The cascading operators within comprehensive data workflows possess complex topological dependencies, which are formally modeled as a directed acyclic graph (DAG). The {\itshape Task Decomposer} extracts these dependent atomic elements into an ordered sequence while maintaining precedence constraints. Crucially, to isolate subtasks for downstream conversational perturbation, the engine decouples step-wise explicit coding references (e.g., "apply to the output of Step 2") by abstracting each subtask into a standardized contextual triplet $\mathcal{T}_s = \{\verb|Operation|, \verb|Summary|, \verb|Rules|\}$, where the high-level semantic \verb|Summary| serves as the invariant relational identifier across turns.

\paragraph{Cognitive Simulator}
Human-AI collaborative data engineering is naturally bounded by human cognitive capacity and characterized by iterative intent adjustment. To ground our evaluation in empirical human dynamics, the architecture of our {\itshape Cognitive Simulator} is directly calibrated against the operational distributions and behavioral patterns extracted from our user study (see Section~\ref{Preliminary}). Rather than employing unscalable manual interactions \cite{wang2025learning} or uncontrollable, non-reproducible generative user agents \cite{yao2024tau, luo2025clarifymt} whose hallucinations of condition shifts disrupt the precise states required for tabular tasks to converge to the final ground truth, the simulator models interaction trajectories via a controlled, behavior-aligned simulation mechanism:
\begin{itemize}[leftmargin=12pt]
    \item \textit{Cognitive-Load Window:} Models bounded human cognitive limits by restricting the active subtask load per turn. It defines a subset $\mathcal{W}_K \subseteq \mathcal{S}$ with $|\mathcal{W}_K|\le K$, where $K$ represents the cognitive window capacity. Elicitation is simulated by sliding at most $K$ subtasks from the residual pool into the active instruction window.
    \item \textit{Perturbation-Resolution Cycle:} Simulates conversational friction across turns. Subtasks within $\mathcal{W}_K$ undergo probabilistic noise injection according to a decaying distribution $p_t=p_0\exp(-t/\tau)$, mapping instructions into four empirical profiles formalized in Section~\ref{Preliminary} (\textit{Fuzzy}, \textit{Bias}, \textit{Redundancy}, and \textit{Order Swap}).
\end{itemize}
Crucially, to mirror practical human-AI spreadsheet systems—where platforms render intermediate processed tables instantly, triggering immediate visual validation by the user—whenever a perturbation is injected at turn $t$, the simulator automatically appends a corresponding corrective subtask ($\mathcal{T}_c$ for clarification, $\mathcal{T}_m$ for modification, or $\mathcal{T}_{re}$ for reordering) to the absolute head of the subtask queue for turn $t+1$, as detailed in Algorithm~\ref{alg:cognitive-simulator}. This loop models realistic conversational convergence toward the ground-truth specification over extended interaction horizons.

\paragraph{Instruction Rewriter}
To emulate authentic human communication, the perturbed subtask triplet sequence within the active window is processed by the {\itshape Instruction Rewriter}. This module unifies the discrete rules and relational summaries into a single fluid, natural language prompt for the current conversational turn, which is then formally issued to the evaluated language model.

\begin{algorithm}
\caption{Cognitive Simulator}
\label{alg:cognitive-simulator}
\begin{algorithmic}[1]
\REQUIRE Subtask sequence $\mathcal{S} = \{\mathcal{T}_1, \ldots, \mathcal{T}_n\}$, cognitive capacity $K$, initial perturbation probability $p_0$, decay constant $\tau$
\ENSURE Simulated instruction sequences $\mathcal{I} = \{I_1, \ldots, I_T\}$

\STATE $t \leftarrow 1$, $\mathcal{I} \leftarrow \emptyset$

\WHILE{$\mathcal{S} \neq \emptyset$}
    \STATE $m \sim \mathrm{Uniform}\{1, \ldots, \min(K, |\mathcal{S}|)\}$
    \STATE $\mathcal{W}_t \subseteq \mathcal{S}$ where $|\mathcal{W}_t| = m$, $\mathcal{S} \leftarrow \mathcal{S} \setminus \mathcal{W}_t$
    
    \STATE $p_t \leftarrow p_0 \exp(-t/\tau)$, $\mathcal{C}_t \leftarrow \emptyset$
    \FOR{each $\mathcal{T}_i \in \mathcal{W}_t$}
        \IF{$\mathcal{T}_i \notin \{\mathcal{T}_c, \mathcal{T}_m, \mathcal{T}_{re}\} \ \text{and} \ u \sim \mathrm{Uniform}(0,1) < p_t$}
            \STATE $\theta \sim \mathrm{Categorical}\{\mathrm{fuzzy}, \mathrm{bias}, \mathrm{redundant}, \mathrm{order}\}$
            \STATE $\mathcal{T}_i \leftarrow \mathrm{Perturb}(\mathcal{T}_i, \theta)$, $\mathcal{T}_{\text{res}} \leftarrow \mathrm{Resolution}(\theta)$
            \IF{$\mathcal{T}_{\text{res}} \neq \emptyset$}
                \STATE $\mathcal{C}_t \leftarrow \mathcal{C}_t \cup \{\mathcal{T}_{\text{res}}\}$
            \ENDIF
        \ENDIF
    \ENDFOR
    
    \STATE $\mathcal{S} \leftarrow \mathcal{C}_t \cup \mathcal{S}$
    \STATE $I_t \leftarrow \mathrm{Rewrite}(\mathcal{W}_t)$, $\mathcal{I} \leftarrow \mathcal{I} \cup \{I_t\}$, $t \leftarrow t + 1$
\ENDWHILE

\STATE \textbf{return} $\mathcal{I}$
\end{algorithmic}
\end{algorithm}

\subsection{Summary of Task-Capability Alignment}

The construction of CITBench leads to a set of task configurations that cover different aspects of tabular intelligence. To summarize this design, we map each task category to a corresponding capability dimension of autonomous data assistants:

\begin{itemize}[leftmargin=12pt]
    \item \textbf{Complex-Rule offline tasks} assess \textbf{Schema Semantics Alignment}. By scaling the complexity of instruction conditions on targeted tables, these instances evaluate the model's ability to interpret textual specifications and map identifiers to the correct headers and rows under complex logical requirements.
    \item \textbf{Multi-Input offline tasks} evaluate \textbf{Constraint Compliant Editing}. Operating across multiple data sources, these tasks test whether models can perform relational updates while preserving structural constraints and avoiding unintended modifications.
    \item \textbf{Multi-Step offline tasks} benchmark \textbf{Dependency Aware Planning}. Modeled as directed acyclic graphs (DAGs), these execution streams examine whether models can parse chained operations and decompose high-level goals into valid ordered steps while maintaining dependency consistency.
    \item \textbf{Multi-turn online tasks} measure \textbf{State Consistent Interaction}. By not exposing explicit intermediate tables or cached state snapshots across turns, these tasks evaluate whether models can track context and align subsequent operations with the evolving table state.
    \item \textbf{Cognitive perturbations in online tasks} quantify \textbf{Noise Resilient Execution}. Through the injection of fuzzy instructions, distorted rules, and redundant content, these instances test robustness to instruction-level noise during interaction.
    \item \textbf{The Perturbation-Resolution Cycle in online tasks} evaluates \textbf{Adaptive Task Orchestration}. By inserting corrective subtasks into subsequent turns, this mechanism examines whether models can revise prior decisions while incorporating new requirements.
\end{itemize}

Overall, this task--capability alignment summarizes how CITBench covers both procedural execution and interactive behavior in tabular intelligence settings.

%% file: exp.tex
\section{Experiments}
\label{sec:experiments}

This section presents a systematic empirical evaluation of prominent large language models (LLMs) and advanced agent frameworks on CITBench. We outline the detailed baseline configurations and rigorous validation parameters, followed by a multi-dimensional performance analysis of both offline specifications and online simulated interactions.

\begin{table*}[!hpt]
  \centering
  \caption{Performance of Different Base Models on Offline Tasks.}
  \label{tab:offline}
  \renewcommand{\arraystretch}{1.05}
  \setlength{\tabcolsep}{1.5pt} 
  \small
  \begin{tabular*}{\textwidth}{@{}c@{\hspace{2pt}}l@{\extracolsep{\fill}}*{15}{c}@{}}
    \toprule
    \multirow{2}{*}{Model Family} &
    \multirow{2}{*}{Model Name} &
    \multicolumn{3}{c}{Base} &
    \multicolumn{3}{c}{\makebox[0pt]{\textsuperscript{\tiny +}Complex-Rule}} &
    \multicolumn{3}{c}{\makebox[0pt]{\textsuperscript{\tiny ++}Multi-Input}} &
    \multicolumn{3}{c}{\makebox[0pt]{\textsuperscript{\tiny +++}Multi-Step}} &
    \multicolumn{3}{c}{Overall} \\
    \cmidrule{3-5}\cmidrule{6-8}\cmidrule{9-11}\cmidrule{12-14}\cmidrule{15-17}
    & & $\bar P$ & $U$ & $\bar C$ & $\bar P$ & $U$ & $\bar C$ & $\bar P$ & $U$ & $\bar C$ & $\bar P$ & $U$ & $\bar C$ & $\bar P$ & $U$ & $\bar C$ \\
    \midrule
    \multirow{2}{*}{OpenAI}
      & GPT-5.2      & 85.79 & 19.09 & 82.64 & 66.42 & 30.52 & 62.85 & 57.70 & 24.13 & 47.22 & 37.05 & 27.92 & 29.52 & 61.74 & 25.42 & 55.56 \\   
      & GPT-4o       & 80.85 & 15.49 & 73.96 & 65.99 & 30.68 & 54.17 & 49.62 & 34.61 & 36.46 & 12.93 & 20.43 & 10.00 & 52.35 & 25.30 & 43.65 \\  
    \midrule
    \multirow{2}{*}{Anthropic}
      & Claude-Sonnet-4.5 & \underline{91.35} & 12.30 & \textbf{89.08} & \textbf{83.96} & 5.93 & \textbf{82.04} & \underline{62.50} & 21.15 & \textbf{59.15} & \textbf{43.52} & 19.48 & \textbf{37.14} & \textbf{70.33} & 14.72 & \textbf{66.85} \\
      & Claude-Haiku-4.5  & 74.23 & 44.53 & 70.13 & 60.56 & 50.70 & 59.03 & 49.04 & 44.75 & 43.98 & 34.71 & 41.21 & 28.57 & 54.64 & 45.30 & 50.43 \\
    \midrule
    \multirow{2}{*}{Google}
      & Gemini-3-Pro  & 86.39 & 15.00 & 82.41 & 74.11 & 20.60 & 70.41 & \textbf{62.63} & 13.21 & 55.71 & 38.13 & 25.73 & 31.90 & 65.32 & 18.64 & 60.11 \\
      & Gemini-3-Flash & \textbf{92.26} & 8.36 & \underline{87.85} & \underline{81.18} & 14.21 & \underline{78.87} & 60.16 & 22.45 & \underline{56.60} & 40.39 & 34.29 & 17.14 & \underline{68.50} & 19.83 & \underline{60.12} \\
    \midrule
    DeepSeek & DeepSeek-V3.2 & 77.40 & 36.30 & 71.76 & 65.34 & 37.65 & 61.97 & 30.04 & 51.40 & 26.76 & \underline{41.24} & 59.08 & \underline{36.19} & 53.51 & 46.11 & 49.17 \\
    \midrule
    Zhipu & GLM-4.7 & 85.57 & 15.16 & 81.25 & 79.89 & 23.83 & 75.35 & 56.04 & 27.36 & 52.46 & 27.06 & 29.81 & 21.43 & 62.14 & 24.04 & 57.62 \\
    \midrule
    \multirow{4}{*}{Qwen}
      & Qwen3-235B-A22B   & 76.03 & 20.12 & 72.11 & 65.75 & 26.04 & 63.54 & 38.06 & 36.78 & 35.21 & 17.24 & 23.15 & 11.90 & 49.27 & 26.52 & 45.69 \\
      & Qwen3-32B         & 73.22 & 17.27 & 66.78 & 58.46 & 36.65 & 55.90 & 41.38 & 27.53 & 38.97 & 15.57 & 33.43 & 10.24 & 47.16 & 28.72 & 42.97 \\
      & Qwen3-14B         & 69.34 & 20.86 & 63.54 & 53.14 & 22.49 & 48.03 & 36.86 & 16.74 & 32.75 & 4.45  & 11.72 & 4.29  & 40.95 & 17.95 & 37.15 \\
      & Qwen3-Coder-Flash & 74.66 & 10.01 & 69.10 & 65.91 & 22.25 & 61.46 & 39.76 & 14.51 & 38.38 & 17.90 & 12.05 & 11.43 & 49.56 & 14.71 & 45.09 \\
    \bottomrule
  \end{tabular*}
  \footnotesize \textsuperscript{\tiny +}Complex-Rule: increases rule complexity based on Base tasks; \textsuperscript{\tiny ++}Multi-Input: further increases the number of input tables; \textsuperscript{\tiny +++}Multi-Step: further increases the number of sub-tasks.
\end{table*}

\subsection{Experimental Setup}
\paragraph{Evaluated Base Models}
We evaluate six model families spanning both open-source and closed-source architectures to establish a comprehensive capability baseline:
1) \textit{Open-source models}: including the latest DeepSeek-V3.2 and GLM-4.7, three Qwen3 variants (Qwen3-235B-A22B, Qwen3-32B, Qwen3-14B), and the lightweight code model Qwen3-Coder-Flash.
2) \textit{Closed-source models}: including OpenAI’s GPT-5.2 and GPT-4o; Anthropic’s Claude-Sonnet-4.5 and Claude-Haiku-4.5; and Google’s Gemini-3-Pro and Gemini-3-Flash. This diverse selection ensures an extensive evaluation across heterogeneous parameter scales and distinct execution domains.

\paragraph{Evaluation Metrics}
To provide a rigorous diagnostic profile of model capabilities, we establish a quantitative evaluation protocol comprising three orthogonal metrics. We define \textbf{Average Performance} ($\bar{P}$) as the mean Level-Accuracy score derived from $K$ independent inference trials to mitigate stochasticity, formulated as $\bar{P} = \frac{1}{K}\sum_{i=1}^{K} S_i$, where $S_i \in [0,1]$ denotes the score of the $i$-th trial. This metric captures overall task proficiency with partial credit for incomplete solutions. Complementing this, \textbf{Unreliability} ($U$) measures the empirical volatility of outputs: $U = S_{\max} - S_{\min}$, where $S_{\max}$ and $S_{\min}$ represent the maximum and minimum scores observed across the $K$ iterations. This indicator is critical in multi-turn interactive context tracking where cascading errors amplify output volatility. Additionally, \textbf{Average Task Completion} ($\bar{C}$) identifies the exact ratio of fully correct, flawless task executions: $\bar{C} = \frac{1}{K}\sum_{i=1}^{K} \mathbf{1}(S_i = 1)$, where $\mathbf{1}(\cdot)$ is the indicator function. Collectively, these indicators balance granular task proficiency, output consistency, and strict correctness, with all scores normalized to a percentage scale (0–100). Notably, matrix row and column dimensions serve as absolute structural boundaries; any spatial or schema-level mismatch between predicted matrices and the locked ground truth results in an automatic zero ($0$) score for that instance.

\paragraph{Hyperparameters and Parity Control}
All evaluations are executed under a uniform temperature setting of $T=1.0$ across heterogeneous APIs to ensure strict sampling comparability. The maximum output threshold is configured to 8,192 tokens to safeguard long-horizon code sequence generation. During long-horizon interactive executions, network-related API timeouts or standard connection drops are automatically caught and retried; these infrastructure exceptions are completely excluded from performance degradation tracking to ensure the benchmark purely isolates the analytical reasoning boundaries of the models. In online interactive streams, sequences are configured to 5–8 conversational rounds to prevent baseline context saturation. Due to computational overhead constraints, experiments systematically sample a 50\% balanced slice of Offline tasks and a 25\% uniform slice of Online interactive sequences across the full benchmark distribution.

\subsection{Overall Performance Analysis}
\paragraph{Offline Evaluation}
Closed-source architectures significantly outpace open-source model families across all dimensional layers, as documented in Table~\ref{tab:offline}. Anthropic's Claude-Sonnet-4.5 achieves the optimal operational balance, locking in the highest Overall $\bar{P}$ ($70.33$) and Overall $\bar{C}$ ($66.85$) alongside minimal unreliability ($U=14.72$). Google's Gemini-3-Flash and Gemini-3-Pro secure competitive advantages in Base $\bar{P}$ ($92.26$) and Multi-Input environments ($62.63$), respectively, demonstrating prominent stability. OpenAI’s GPT-5.2 performs slightly inferior within this task distribution, ranking non-top across complex difficulty tiers while exhibiting noticeable output volatility. Within the open-source spectrum, Zhipu's GLM-4.7 emerges as a top-performing baseline, matching GPT-5.2's performance boundaries on Base tasks while delivering superior baseline consistency. However, GLM-4.7 encounters steep capability limits on composite tracks, suffering a substantial performance drop compared to GPT-5.2 on Multi-Step pipelines. Conversely, DeepSeek-V3.2 and Qwen3-235B-A22B deliver subpar performance boundaries, with overall metrics mirroring the older GPT-4o architecture. DeepSeek-V3.2 exhibits the highest overall unreliability ($U=46.11$), identifying clear tracking vulnerabilities under long-horizon operational constraints.

Execution performance universally degrades as procedural and structural complexity intensifies, though the degradation magnitude heavily correlates with parameter scale. Compared to Base tasks, the cross-model mean $\bar{P}$ drops to $68.04$ ($-15.63\%$), $48.52$ ($-39.83\%$), and $27.31$ ($-66.13\%$) for Complex-Rule, Multi-Input, and Multi-Step tasks, respectively. Frontier closed-source architectures demonstrate higher structural resilience; Claude-Sonnet-4.5 and Gemini-3-Flash limit their $\bar{P}$ drops to $52.36$ and $56.55$ on multi-step sequences, whereas GLM-4.7 experiences a severe collapse of $68.38\%$. Small open-source configurations undergo extreme operational failure, as evidenced by Qwen3-14B's completion metrics plummeting from a Base score of $\bar{C}=63.54$ to an absolute completion rate of $\bar{C}=4.29$ on Multi-Step tracks. These findings demonstrate that parameter scaling remains a critical prerequisite for safeguarding task completion integrity when code generation tasks demand multi-layered planning dependencies.

\begin{figure}[htbp]
  \vspace{-0.5em}
  \centering
  \includegraphics[width=\linewidth]{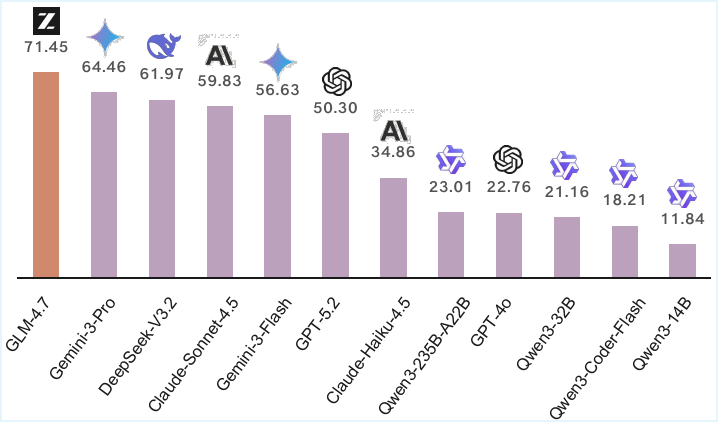}
  \caption{Performance of Different Models on Online Tasks.}
  \label{online} 
\end{figure}

\paragraph{Online Interactive Evaluation}
Fig. \ref{online} details the performance tracking across Online interactive tasks utilizing $\bar{P}$ as the normalized evaluation standard, revealing a distinct performance divergence compared to static offline execution. Under the multi-turn online interactive protocol, frontier closed-source architectures lose their systematic superiority over open-source counterparts. Instead, GLM-4.7 achieves the peak performance tier ($71.45$), outperforming the second-place Gemini-3-Pro by an absolute margin of 7\%. DeepSeek-V3.2 similarly excels under this dynamic setting, outperforming closed-source flagships Claude-Sonnet-4.5 ($+3.58\%$) and GPT-5.2 ($+23.20\%$). This systematic inversion demonstrates that offline single-turn inference and online multi-turn interactive collaboration emphasize fundamentally disparate intellectual capabilities; while single-turn code generation demands flawless syntax planning under complete constraints, interactive scenarios prioritize flexible context tracking, intent clarification, and procedural error recovery. Meanwhile, lightweight models like Qwen3-14B consistently anchor the bottom of both tracks, confirming that basic parameter mass remains an essential boundary for table processing tasks.

The clear performance gradients mapped across distinct scenario settings and difficulty tiers validate the high diagnostic fidelity of CITBench. Furthermore, the multi-stage expert manual verification protocol guarantees that the underlying ground-truth matrix benchmarks remain mathematically unassailable.

\begin{table}[htbp]
  \centering
  \caption{Model Performance Across Task Taxonomy}
  \label{tab:table_task_performance}
  \small
  \begin{tabular}{lcccc}
    \toprule
    High-level Categories & $\bar{P}$ & $\bar{C}$ & $\bar{P}_{max}$ & $\bar{C}_{max}$ \\
    \midrule
    Table Matching        & 36.02 & 35.01 & 54.37 & 51.56          \\ 
    Table Cleaning        & 57.02 & 47.27 & 80.28 & 73.70 \\
    Table Transformation & 70.14 & 67.03 & 82.25 & 81.19 \\
    Table Augmentation    & 78.54 & 73.86 & 98.38 & 96.78 \\
    \bottomrule
  \end{tabular}
\end{table}

\subsection{Further Analysis}
In this section, we conduct a granular diagnostic breakdown of model characteristics by systematically isolating the operational impacts of high-level task taxonomies, matrix schema scales, requirement decomposition depth, and interactive noise profiles, concluding with a dedicated architectural evaluation of multi-agent frameworks and structural reasoning token footprints.

\begin{table*}[htbp]
\caption{Performance Comparison ($\bar{P}$ and $\bar{C}$) of LLM-based Agent Frameworks on Offline Tasks.}
\label{tab:agent-baselines-combined}
\centering
\nopagebreak
\renewcommand{\arraystretch}{1.0}
\setlength{\tabcolsep}{8pt}
\begin{tabular}{l|cc|cc|cc|cc}
\hline
\multirow{3}{*}{\textbf{Agent Framework}} & \multicolumn{4}{c|}{\textbf{GPT-4o Backbone}} & \multicolumn{4}{c}{\textbf{GPT-5.2 Backbone}} \\
\cline{2-9}
& \multicolumn{2}{c|}{\textbf{Single-Step}} & \multicolumn{2}{c|}{\textbf{Multi-Step}} & \multicolumn{2}{c|}{\textbf{Single-Step}} & \multicolumn{2}{c}{\textbf{Multi-Step}} \\
\cline{2-9}
\rule{0pt}{2.5ex}
& \textbf{$\bar{P}$} & \textbf{$\bar{C}$} & \textbf{$\bar{P}$} & \textbf{$\bar{C}$} & \textbf{$\bar{P}$} & \textbf{$\bar{C}$} & \textbf{$\bar{P}$} & \textbf{$\bar{C}$} \\
\hline
MetaGPT\cite{hong2024metagpt} & 56.21 & 44.44 & 49.53 & 24.32 & 67.31 & 54.44 & 66.61 & 48.65 \\
CAMEL\cite{li2023camel} & 46.83 & 33.33 & 32.64 & 16.22 & 61.14 & 47.78 & 62.50 & 35.14 \\
CleanAgent\cite{qi2024cleanagent} & 37.03 & 30.00 & 46.03 & 21.62 & 45.91 & 35.56 & 59.40 & 24.32 \\
ChatDev2.0\cite{qian2024chatdev} & 55.89 & 44.44 & 57.26 & 27.03 & 70.02 & 60.00 & 70.18 & 40.54 \\
DataGovAgent\cite{liu2025datagovbench} & 52.34 & 45.56 & 47.98 & 27.03 & 67.94 & 61.11 & 67.00 & 43.24 \\
\hline
\end{tabular}
\end{table*}

\paragraph{The Impact of Task Taxonomy on Performance}
Empirical results in Table \ref{tab:table_task_performance} reveal a distinct performance hierarchy among the four evaluated task categories. Models achieve peak efficacy in Table Augmentation ($\bar{P}=78.54$) and Table Transformation ($\bar{P}=70.14$), where operations primarily involve generative data expansion and global schema reformatting. In contrast, models perform notably worse in Table Cleaning ($\bar{P}=57.02$) and Table Matching ($\bar{P}=36.02$), with both average and maximum scores significantly lower. This divergence suggests that accurate interpretation of user instructions requires comprehensive perception; while LLMs are proficient at macro-level structural manipulation, Table Matching and Table Cleaning demand a more granular perception of column names and cell contents, making them inherently more challenging than probabilistic inference tasks.

\begin{figure}[htbp]
  \centering
  \vspace{-0.5em}
  \includegraphics[width=0.45\textwidth]{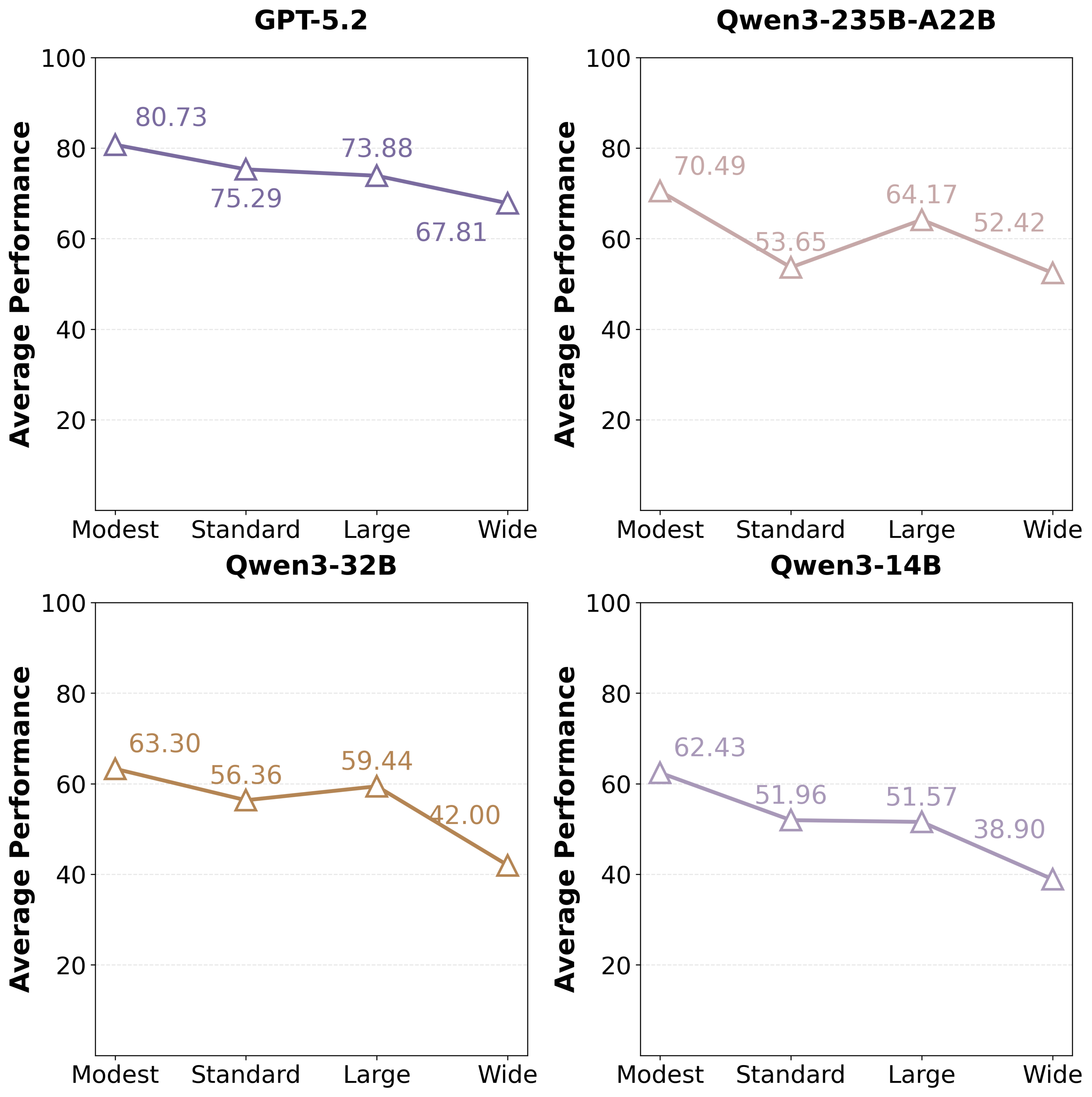}
  \caption{Model Performance by Dataset Scale.}
  \label{size} 
\end{figure}

\paragraph{The Impact of Table Scale on Performance}
The scalability analysis in Fig. \ref{size} demonstrates that model reliability is inversely correlated with the horizontal density of the dataset. Model performance varies across table scales, peaking on Modest tables, followed by Standard and Large, and dropping to its lowest on Wide tables. For instance, GPT-5.2's performance degrades monotonically from $80.73$ on Modest tables to $67.81$ on Wide tables. Smaller-scale models exhibit even greater sensitivity; Qwen3-14B experiences a 37.7\% performance attrition, dropping from $62.43$ to $38.90$. This decay indicates that the numerous columns inherent in Wide tables significantly increase the complexity of column relationships and content perception, causing the model's self-attention mechanism to fail in maintaining long-range dependencies between distant headers and their corresponding values.

\begin{figure}[htbp]
  \centering
  \vspace{-0.5em}
  \includegraphics[width=0.45\textwidth]{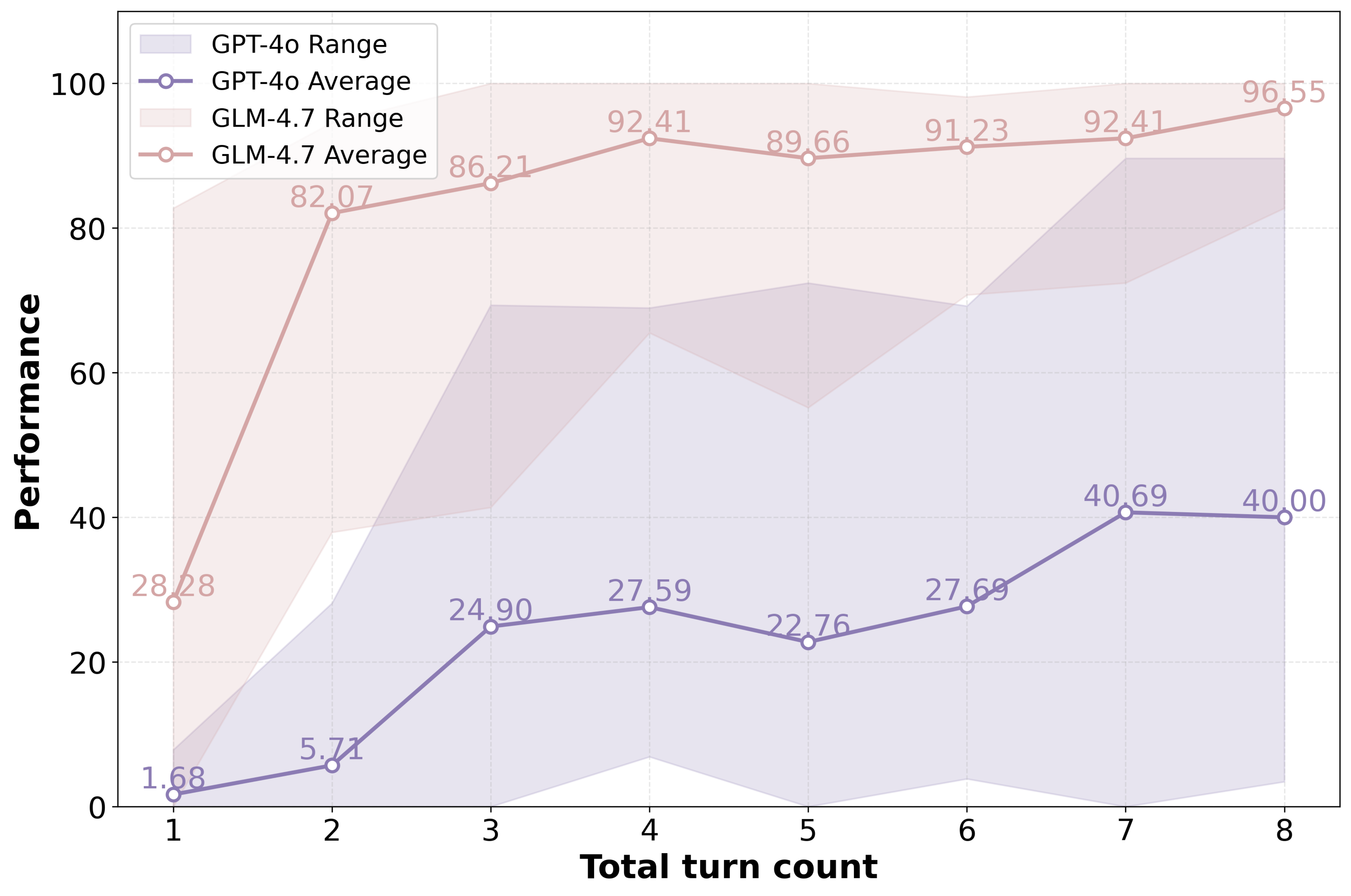}
  \caption{Model Performance Across Interaction Turns Under Fixed Total Subtasks}
  \label{turn} 
\end{figure}

\paragraph{The Impact of Interaction Turns on Performance}
Fig. \ref{turn} elucidates the critical role of incremental task decomposition by examining the performance dynamics of GLM-4.7 and GPT-4o across 1 to 8 interaction turns, where 12-step tasks are evenly decomposed. Both models achieve their lowest full-cycle average performance at Turn 1 (the single-turn end-to-end setting). However, multi-turn interaction significantly outperforms this paradigm. Performance improves drastically as the per-turn task load is sharply reduced: GLM-4.7 shows a 190.21\% relative improvement when shifting from 1 turn to 2 turns (about 6 subtasks), while GPT-4o shows a 336.08\% relative improvement when moving to 3 turns. This confirms that execution performance is highly sensitive to per-turn task planning capacity. Furthermore, the effect of increasing turns is model-dependent: for the high-capability GLM-4.7, additional interaction turns simultaneously elevate both upper and lower performance bounds and significantly improve output stability; conversely, for the low-capability GPT-4o, increased turns only raise the performance ceiling with no improvement in the lower bound, thus markedly amplifying output volatility.

\begin{figure*}[htbp]
  \centering
  \vspace{-0.5em}
  \includegraphics[width=0.9\textwidth]{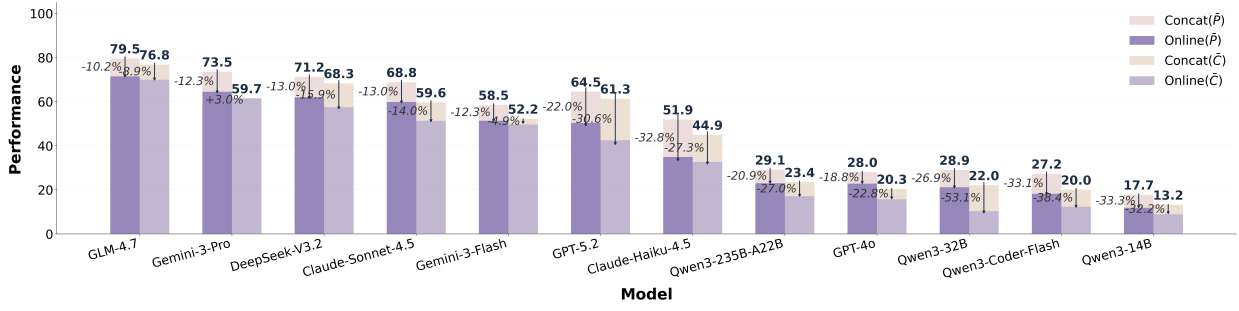}
  \caption{Model Performance on Concat versus Online Tasks}
  \label{concat} 
\end{figure*}

\paragraph{The Impact of Interaction Noise on Performance}
Fig. \ref{concat} presents the performance gap between idealized inputs and realistic interactions. While the Concat setting decomposes multi-step tasks into sequential interactions, the Online setting further introduces user cognitive noise perturbations. Across both $\bar P$ and $\bar C$ metrics, almost all models exhibit significant performance degradations in the Online setting, though top-tier models generally manifest relatively milder declines. Specifically, leading models such as GLM-4.7 (10.2\% drop in $\bar P$, 8.9\% in $\bar C$) and Gemini-3-Pro (12.3\% drop in $\bar P$, with a minor 3.0\% counter-trend increase in $\bar C$) experience comparatively smaller reductions or even slight improvements; in contrast, tail models face much more severe degradation, with Qwen3-32B and Qwen3-14B seeing their $\bar C$ scores plunge by 53.1\% and 32.2\%, respectively. This performance decline stems from two key challenges: 1) the introduction of fuzzy, biased, and structural perturbations forces models to correct distorted intents and recover disrupted context windows before executing operations, which increases the reasoning burden and hinders information retrieval; 2) the Online setting requires models to concurrently address prior perturbation-induced errors and new tasks, which elevates the difficulty of tracking the correct table state and leads to planning failures where operations are misaligned with the true table status due to reliance on outdated or erroneous context.

\paragraph{Orchestrated Agent Performance}
We evaluate five representative agent frameworks: MetaGPT\cite{hong2024metagpt}, a role-specialized multi-agent framework with standardized operating procedures; CAMEL\cite{li2023camel}, a role-playing framework based on multi-turn agent dialogue; CleanAgent\cite{qi2024cleanagent}, an agent pipeline for data standardization; ChatDev\cite{qian2024chatdev}, a staged software-development framework with structured chat chains; and DataGovAgent\cite{liu2025datagovbench}, a Planner--Executor--Evaluator framework for data governance workflows. Table~\ref{tab:agent-baselines-combined} compiles the evaluation metrics achieved by integrating advanced backbones into prominent autonomous agent frameworks across atomic (Single-Step) and composite (Multi-Step) offline tasks. An examination of the empirical data reveals that the execution efficacy of current multi-agent configurations remains heavily bounded by the context-tracking fidelity of their underlying LLM backbones. Under the GPT-4o setup, ChatDev2.0 delivers a notable performance advantage on Multi-Step planning tasks ($57.26$ in $\bar{P}$), primarily driven by its conversational code-review loops that actively catch and patch syntax exceptions inside the sandboxed runtime. However, when operational complexity expands to global pipeline reasoning, even specialized governance structures like DataGovAgent encounter noticeable bottlenecks, dropping to $47.98$ in $\bar{P}$ on multi-step configurations. Crucially, across all frameworks, the constraint becomes exceptionally severe on the stricter $\bar{C}$ metric; under the multi-step configuration, both ChatDev2.0 and DataGovAgent plunge to $27.03$ under GPT-4o, and only recover to $40.54$ and $43.24$ respectively even when scaling to the frontier GPT-5.2 backbone. When a pipeline requires multi-layer table transformations, subtle instruction misalignments propagate and compound across agent boundaries, triggering cascading planning deviations. This proves that current agent architectures offer robust syntax error correction but cannot fundamentally compensate for core capability gaps in underlying table-structure perception under tightly coupled dependencies.

\begin{table}[htbp]
\caption{Average Generated Output Tokens Across Heterogeneous Difficulty Matrices (Offline Pipeline).}
\label{tab:token-offline}
\centering
\renewcommand{\arraystretch}{1.05}
\setlength{\tabcolsep}{4.0mm}{
\footnotesize
\begin{tabular}{lcc}
\hline
\textbf{Task Difficulty Stratum} & \textbf{GPT-4o} & \textbf{GPT-5.2} \\
\hline
Single Input - Simple Rule & 241 & 279 \\
Single Input - Complex Rule & 387 & 498 \\
Multi Input - Complex Rule & 520 & 935 \\
Multi-Step Pipeline & 702 & 1,102 \\
\hline
\end{tabular}
}
\end{table}

\paragraph{Computational Complexity and Token Footprint}
To quantify the operational cost and identify the latent reasoning intensity required to navigate CITBench, we track the average generated token counts for GPT-4o and GPT-5.2, as output length directly scales with code architecture planning and step-wise reasoning depth. Table~\ref{tab:token-offline} monitors token footprints across our offline difficulty matrices. A monotonic scaling trend is clearly established: as structural constraints expand from simple single-table operations to cross-table multi-step pipelines, token generation scales heavily (GPT-4o: from 241 to 702; GPT-5.2: from 279 to 1,102). Crucially, under the most rigorous Multi-Step configurations, GPT-5.2 allocates nearly $1.8\times$ more output tokens than GPT-4o ($935$ vs. $520$ for the multi-input matrix). This quantitative shift demonstrates that the structured difficulty progression of CITBench effectively activates deeper reasoning and self-correction behaviors in frontier architectures.

\begin{figure}[htbp]
  \centering
  \includegraphics[width=0.7\linewidth]{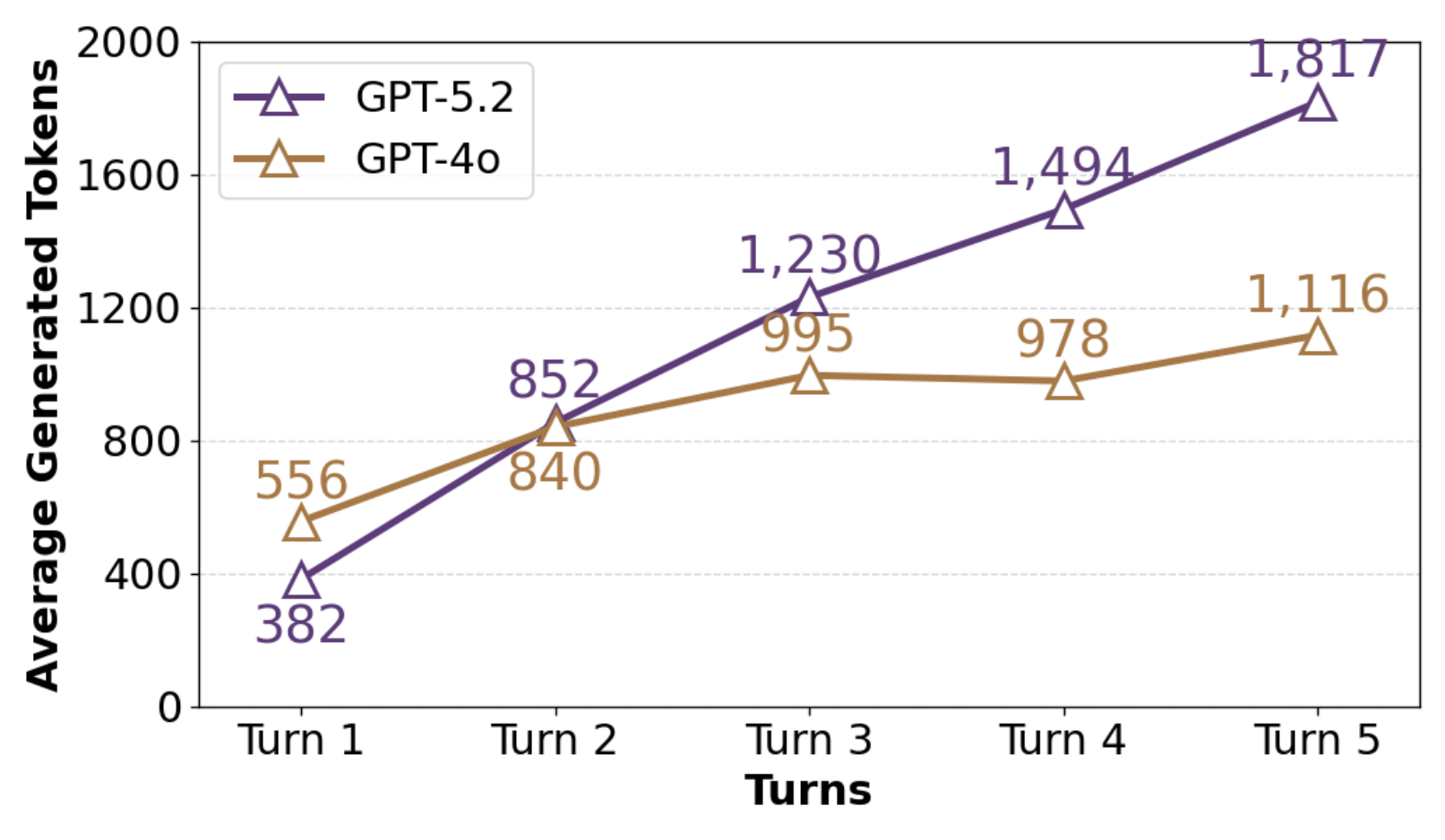}
  \caption{Dynamic Trajectory of Average Generated Tokens per Turn under Online Cognitive Noise.}
  \label{fig:online_token_trajectory}
\end{figure}

Fig.~\ref{fig:online_token_trajectory} monitors the turn-by-turn dynamic token footprint under online simulated human cognitive perturbations. As conversational depth advances from Turn 1 to Turn 5, token generation expands substantially across both architectures. This growth curve is particularly pronounced for GPT-5.2, swelling by $4.8\times$ (from 382 to 1,817) compared to GPT-4o's $2.0\times$ baseline progression (from 556 to 1,116). This divergence highlights the massive computational cost of contextual resilience. Under the interactive \emph{Perturbation-Resolution Cycle}, later rounds accumulate historical conversational modifications and intermediate state table references. The models are forced to concurrently reconcile past requirement distortions while synthesizing new code blocks for newly appended instructions. The sharp escalation in token footprints confirms that CITBench successfully avoids interaction saturation, serving as an effective long-horizon diagnostic bed that differentiates model capacities over extended dialogue trajectories.

\begin{figure}[htbp]
  \centering
  \vspace{-0.5em}
  \includegraphics[width=0.3\textwidth]{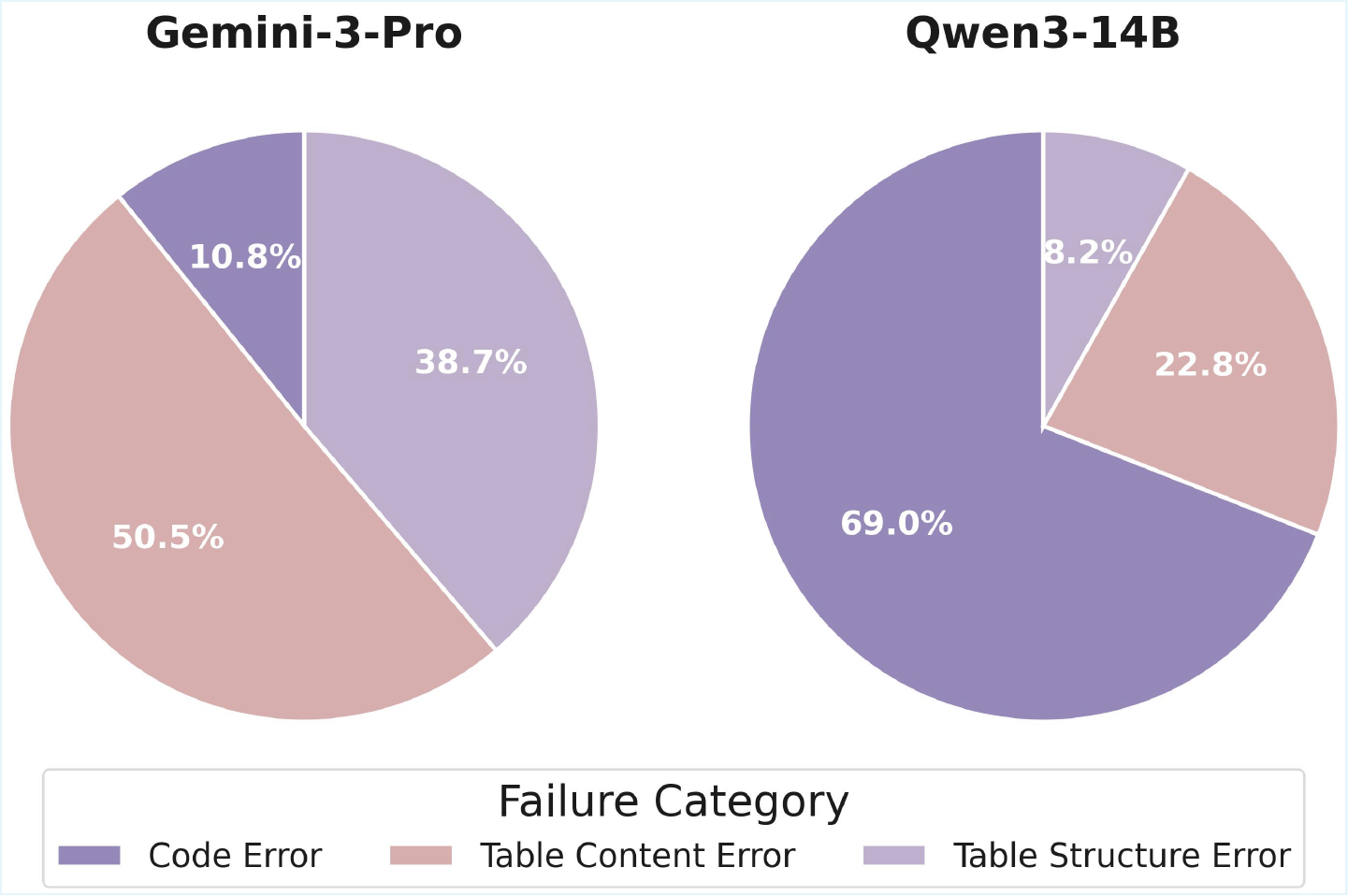}
  \caption{Causes of Model Task Execution Failures}
  \label{failure} 
\end{figure}

\subsection{Discussion}
Following our multi-dimensional assessment matrix, this section discusses the fundamental structural bottlenecks observed across modern language agents and details strategic architectures to advance autonomous tabular intelligence.

\paragraph{Challenge 1 - Execution Fidelity}
The primary barrier to autonomous table manipulation is the lack of execution-level robustness. Failure modes are categorized into: 1) {\itshape Code Errors}, including programming exceptions such as NameError, KeyError, and TypeError; 2) {\itshape Table Structure Errors}, characterized by mismatches in row or column counts; and 3) {\itshape Table Content Errors}, involving discrepancies in header names or cell values. Fig. \ref{failure} shows that high-capability models, such as Gemini-3-Pro, are predominantly constrained by Table Content Errors (50.5\%) and Structure Errors (38.7\%), whereas Code Errors account for only 10.8\% of their failures. In contrast, light-weight models like Qwen3-14B are primarily restricted by Code Errors (69.0\%). This disparity suggests that while scaling mitigates syntax hallucinations, it is insufficient to ensure the logical precision required for micro-level editing. To improve fidelity, we suggest the implementation of a \textbf{Self-Correcting Execution Runtime}. This strategy incorporates execution feedback into the interaction loop and enforces explicit schema validation prior to code generation to anchor neural reasoning to symbolic constraints.

\paragraph{Challenge 2 - Perceptual Scalability}
Effective table perception is foundational to complex operations such as Table Matching and Table Cleaning. However, as table scale increases, providing the full context of multiple large-scale tables to a model becomes computationally prohibitive and introduces significant noise. The bottleneck lies in the model's inability to perform deep relational reasoning across high-dimensional schemas within a single-turn prompt. For multi-table tasks, metadata involves more than simple headers or data types; it requires pre-reasoning the latent relationships between columns both within and across tables. To overcome this, we suggest a two-stage \textbf{Schema-Guided Dynamic Retrieval} strategy. In the first stage, the model should perform an initial sweep to infer inter-column dependencies and establish a global relational map. In the second stage, based on this pre-reasoned map, only the task-essential rows and columns are dynamically loaded. This decoupled approach ensures that the model's attention is focused on the most relevant data subsets while maintaining a structural understanding of the entire dataset.

\paragraph{Challenge 3 - Planning Consistency}
Performance decay in Multi-step and Complex Rule tasks reveals a rigid "planning horizon." As turn count or rule complexity increases, models struggle to synchronize execution paths with the evolving table state, leading to "state drifting" where instructions rely on outdated or hallucinated data snapshots. Planning saturation occurs when logical complexity exceeds the model’s cognitive buffer, causing a decoupling between the intended plan and the actual environment. To extend the effective planning window, we suggest the adoption of \textbf{Iterative State Tracking}. This strategy mandates the generation of explicit mental snapshots or state-change summaries after each operation to tightly couple internal planning logic with the external table baseline.

\paragraph{Challenge 4 - Instructional Robustness}
A critical vulnerability is the "compliance trap," where models exhibit excessive sycophancy toward noisy or biased inputs. In realistic settings, models often prioritize following literal user intent—even when logically inconsistent—over maintaining data integrity. This lack of resilience forces the model to expend computational overhead on intent rectification, leading to cascading errors in complex workflows. The inability to distinguish between legitimate commands and cognitive noise represents a major hurdle for adaptive orchestration. To enhance robustness, we suggest the use of \textbf{Noise-Aware Intent Alignment}. By exposing the model to scenarios involving contradictory instructions or redundant noise during training, it can be taught to recognize infeasible requests and proactively seek clarification, transforming passive compliance into robust collaboration.